\documentclass[preprint,11pt]{elsarticle}

\usepackage{booktabs}
\usepackage{array}
\usepackage{makecell}   
\usepackage{amsmath}
\usepackage{xcolor}
\usepackage{url}
\usepackage{microtype}  
\usepackage{adjustbox}

\newenvironment{fittable}
  {\begin{adjustbox}{max width=\linewidth}}
  {\end{adjustbox}}

\newcommand{\code}[1]{\texttt{\small #1}}

\usepackage[hidelinks]{hyperref}   
\usepackage{xurl}                  

\newenvironment{wtable}{\begin{table}[htbp]}{\end{table}}
\newenvironment{ntable}{\begin{table}[htbp]}{\end{table}}

\journal{}
\makeatletter
\def\ps@pprintTitle{%
  \let\@oddhead\@empty
  \let\@evenhead\@empty
  \def\@oddfoot{\footnotesize\itshape\hfill\today}%
  \let\@evenfoot\@oddfoot}
\makeatother

\begin{document}

\begin{frontmatter}

\title{Provenance, Not Behaviour: A Serialisation Artifact in Edge-IIoTset and
a Leakage-Free Benchmark for Precision-Agriculture Intrusion Detection}

\author[inst1]{Mostafa M. Galal}
\ead{mostafam.galal82@gmail.com}

\affiliation[inst1]{organization={Department of Computer and Systems
                                  Engineering, Alexandria University},
                    city={Alexandria},
                    country={Egypt}}

\begin{abstract}
Edge-IIoTset is the reference benchmark for machine-learning intrusion
detection in the industrial Internet of Things, and results reported on it
cluster above 99\%. We show that much of that performance is not intrusion
detection. The preprocessing recipe distributed with the dataset instructs
researchers to one-hot encode seven categorical columns. Four of them separate
attack from normal traffic with an accuracy of 1.0000 on their own, through the
spelling of the placeholder written for an absent protocol field: the string
\code{0} in the normal-traffic branch of the dataset build against \code{0.0}
in the attack branch. The label is recoverable from a serialisation artifact
encoding file provenance, with no network behaviour modelled, and separates
every row of both curated subsets. Under 5-fold $\times$ 3-repeat
cross-validation, five of six standard classifiers attain exactly $1.0000 \pm
0.0000$ accuracy and the sixth attains 0.99998. Under a corrected protocol,
naive Bayes falls by 0.3005 macro-F1 and the strongest model settles at
$0.9503 \pm 0.0011$. Label, ordinal and frequency encoding leak identically.
Because the curated subsets also lack Modbus and per-device identity, we
rebuild the benchmark from the raw captures under uniform parsing, producing
\textbf{AgriEdge}: 1{,}276{,}122 rows, five devices with full attribution, and
no column separating the classes above 0.0288. A leave-one-device-out sweep
locates the generalisation boundary at the perception/actuation layer, where
random forest falls from 0.9988 to 0.5083 balanced accuracy. Non-IID federated
partitioning costs at most 0.0037 macro-F1, but a 20-round LoRaWAN training run
costs 4.6 hours of uplink.

\end{abstract}


\end{frontmatter}

\section{Introduction}
\label{sec:intro}

Precision agriculture has become one of the more consequential deployments of
the industrial Internet of Things (IIoT). Soil-moisture probes, pH sensors,
water-level floats and climate stations feed telemetry to edge gateways over
lightweight publish/subscribe protocols, and those gateways in turn actuate
irrigation pumps and valves over industrial fieldbus protocols such as Modbus
TCP. The consequences of compromise are physical and immediate: a falsified
soil-moisture reading can drive an irrigation controller to flood a field, and
a denial-of-service attack against a message broker can prevent an irrigation
trigger from firing during the narrow window in which it matters. Unlike an
enterprise network, a farm has no on-site security staff, intermittent
connectivity, and gateways that share a power budget with the machinery they
control.

These constraints have made machine-learning intrusion detection an attractive
proposition, and the Edge-IIoTset dataset of Ferrag et
al.~\cite{ferrag2022edgeiiotset} has become the dominant vehicle for evaluating
it. The dataset is unusually well-suited to the task on paper: it was captured
from a physical seven-layer testbed with real sensors, it spans fourteen attack
classes across MQTT, Modbus, HTTP and DNS, and it is explicitly partitioned to
support both centralised and federated learning. It is also convenient, at
1.2~GB in its curated form, and it ships with a \code{Readme.txt} giving a
step-by-step preprocessing recipe.

The results reported on it are, uniformly, extraordinary. Recent work reports
99.27\% accuracy with 99.21\% F1~\cite{discover2026hybrid} and 99.94\%
accuracy~\cite{hasan2025autoencoder}. Such numbers are
ordinarily read as evidence that the problem is close to solved and that
research attention should move to deployment concerns---model compression,
federated aggregation, inference latency.

This paper argues that the numbers should instead be read as a warning. We
began with the intention of building a precision-agriculture intrusion
detection system on Edge-IIoTset, following the standard pipeline. Before
training, we audited the features the dataset's own documentation instructs
researchers to construct. What we found is that the recipe manufactures a
near-perfect label proxy out of a serialisation artifact.

\subsection{The artifact in brief}

Edge-IIoTset encodes an absent protocol field as a zero rather than as a
null---reasonable, since a packet carrying no MQTT layer has no MQTT topic. The
normal-traffic captures and the attack captures were, however, parsed
separately and concatenated afterwards, and the two branches serialised that
zero differently: as the string \code{0} in one and as the string \code{0.0} in
the other. Step~5 of the distributed recipe then applies
\code{pandas.get\_dummies} to seven columns containing these placeholders.
Dummy encoding treats \code{0} and \code{0.0} as distinct tokens, so the
resulting binary features encode \emph{which file a row came from}. Because
normal traffic and attack traffic came from different files, file provenance is
the label.

The consequence is that four of the seven columns in the recipe recover the
binary label with an accuracy of 1.0000 in isolation. No model that consumes
those features is doing intrusion detection; it is reading a build artifact.

\subsection{Contributions}

\begin{enumerate}
\item \textbf{We identify and characterise a label-leakage mechanism in the
canonical Edge-IIoTset preprocessing recipe}, localise it to placeholder
serialisation, and quantify it with three independent measures (token purity,
single-column held-out accuracy, normalised mutual information). Four columns
reach a separation rate of 1.0000 (Section~\ref{sec:audit}).

\item \textbf{We measure the artifact's contribution to reported performance}
under 5-fold $\times$ 3-repeat cross-validation. Five of six standard
classifiers reach exactly $1.0000 \pm 0.0000$ accuracy in every fold and the
sixth reaches 0.99998; under the corrected protocol the same models span 0.6995
to 0.9503 macro-F1. A 1D-CNN and a deep MLP,
evaluated separately on a single stratified split, reach 1.0000 accuracy under
the same recipe and lose more than 0.21 macro-F1 under correction
(Section~\ref{sec:worth}).

\item \textbf{We show the leak is encoding-agnostic.} One-hot, label, ordinal
and frequency encoding all yield identical 1.0000 single-column accuracy,
because each is a bijection on the token set. The exposed population is
therefore every study that treated these columns as categorical, not only those
that copied the distributed recipe (Section~\ref{sec:encoding}).

\item \textbf{We show the curated subsets cannot support agricultural
research}, with Modbus effectively absent and device identity stripped, and we
rebuild the benchmark from the raw captures under uniform parsing
(Section~\ref{sec:cannot}, Section~\ref{sec:agriedge}).

\item \textbf{We locate the generalisation boundary in agricultural IIoT
detection.} A leave-one-device-out sweep over all five devices shows that
transfer between perception sensors is near-free---every model but naive Bayes
scores 0.9880--0.9999 balanced accuracy when soil moisture, water level or pH
is withheld, and 0.9106--0.9385 for temperature and humidity---while transfer
across the perception/actuation layer collapses to 0.4877--0.5083, straddling
the trivial baseline, for two of the six (Section~\ref{sec:lodo}).

\item \textbf{We show that the federated literature is optimising the wrong
constraint.} Non-IID partitioning costs at most 0.0037 macro-F1, while training
a 28{,}450-parameter model over LoRaWAN costs 4.6 hours of uplink
(Section~\ref{sec:fed}).

\item \textbf{We characterise edge deployment cost}, finding a
342$\times$ inference-latency spread and a 171$\times$ model-size spread across
models whose random-split accuracies are indistinguishable
(Section~\ref{sec:edge}).

\item \textbf{We release the complete pipeline}---audit tooling, benchmark
construction, and experiments---so that the audit can be applied to other
datasets (Section~\ref{sec:repro}).\footnote{\url{https://github.com/MostafaGalal1/agriedge}}
\end{enumerate}

\section{Background}

\subsection{Precision agriculture as a layered IIoT system}

The precision-agriculture deployments that Edge-IIoTset's testbed models
decompose into three layers, and the security properties of each differ
sharply.

\textbf{The perception layer} comprises the sensors that observe the physical
field: soil moisture, pH, water level, temperature and humidity. Their traffic
is low-rate, highly periodic, and semantically narrow---a soil-moisture probe
emits a scalar at a fixed interval. This regularity is what makes anomaly
detection tractable at this layer, and also what makes the layer a target: an
adversary who can alter a reading controls the actuation logic downstream
without ever touching the actuator.

\textbf{The network layer} carries that telemetry. MQTT dominates sensor
reporting because its publish/subscribe model and small header suit constrained
devices, while Modbus TCP remains the lingua franca for actuation because it is
what irrigation controllers and PLCs speak. Neither protocol authenticates or
encrypts by default in typical farm deployments. A man-in-the-middle positioned
between a probe and the broker can rewrite a moisture reading; a
denial-of-service attack against the broker prevents irrigation triggers from
firing at all.

\textbf{The edge system layer} performs local processing. Agricultural sites
are often served by cellular or LoRaWAN backhaul with severely constrained
uplink, so detection must run on the gateway rather than in the cloud. This is
the layer at which the tension between detection quality and resource budget
binds.

\subsection{The Edge-IIoTset dataset}

Edge-IIoTset~\cite{ferrag2022edgeiiotset} was captured from a physical testbed
spanning IoT sensors, edge gateways, SCADA components and adversary hosts. It
is distributed in three parts: raw per-device normal-traffic captures (CSV and
PCAP), raw per-attack captures, and two curated subsets---
\code{ML-EdgeIIoT-dataset.csv} (157{,}800 rows) and
\code{DNN-EdgeIIoT-dataset.csv} (2{,}219{,}201 rows)---intended for traditional
machine learning and deep learning respectively. Both curated subsets carry 63
columns: 61 protocol features, a binary \code{Attack\_label}, and a 15-class
\code{Attack\_type}.

Nearly all published work uses the curated subsets, and follows the
\code{Readme.txt} recipe: drop fifteen identifier and payload columns, drop
nulls and duplicates, and dummy-encode seven categorical columns. That recipe
is the object of our audit.

\subsection{Related work}

\paragraph{Performance on Edge-IIoTset} The literature is dense and its results
are tightly clustered near ceiling. Reported figures include 99.27\% accuracy
with 99.21\% F1 across four benchmark datasets~\cite{discover2026hybrid} and
99.94\% accuracy with a confusion matrix containing no errors of either
kind~\cite{hasan2025autoencoder}.
Architectures range from autoencoder-based lightweight feature
learning~\cite{hasan2025autoencoder} to hybrid CNN-DNN models and TinyML
deployments for energy-aware edge inference~\cite{screports2026tinyml}.

\paragraph{Emerging scepticism} A parallel literature has begun questioning
whether these numbers mean what they appear to. Hakim et
al.~\cite{hakim2026crossdomain} train four lightweight architectures on one
IIoT dataset and evaluate them, without retraining, on two structurally
distinct ones; they find that models rely on coarse port-category shortcuts
appearing 96--435$\times$ more frequently in source-domain attacks, and that
prior work evaluates models only within their training network, leaving
behaviour on unseen networks unverified. A dataset-centric review in
\emph{Frontiers in Big Data}~\cite{frontiers2026datasetcentric} catalogues
evaluation biases across IoT/IIoT intrusion detection, noting that published
results can contain undocumented preprocessing operations, hidden leakage
mechanisms, or inconsistent validation protocols, and that no studies report
$k$-fold cross-validation on Edge-IIoTset, relying instead on 80/20 splits.
Recent work on cross-domain heterogeneous
benchmarking~\cite{bridge2026tchnet} and dataset-centric evaluation of
federated intrusion detection~\cite{screports2026federated} points in the same
direction.

\paragraph{Benchmark integrity as a research programme} A distinct line of work
audits intrusion detection benchmarks directly rather than proposing models for
them. Engelen et al.~\cite{engelen2021cicids} reverse-engineer CICIDS2017 and
find labelling and traffic-generation faults severe enough to change reported
conclusions. Flood et al.~\cite{flood2024designsmells} generalise this into a
catalogue of six recurring \emph{data design smells} obtained by manual
analysis of seven highly-cited benchmark datasets. Two of the six sit near our
finding: \emph{highly dependent features}, covering features of outsized
importance that are unrelated to the underlying mechanism of an attack, and
\emph{wrong label}, covering mislabelled data produced by inaccurate ground
truth in a generation testbed. Arp et al.~\cite{arp2022dosdonts} survey thirty top-tier
security papers and codify the recurring methodological pitfalls of
learning-based security, spurious correlation among them.

This paper sits inside that programme rather than alongside it, and two things
distinguish its contribution. First, Edge-IIoTset is not among the seven
datasets Flood et al. examine, and we are not aware of any comparable audit of
it; the catalogue therefore does not yet cover the dataset that IIoT and
agricultural work overwhelmingly relies on. Second, the fault sits where the
catalogue does not look. All six smells locate the problem in the testbed---in
how traffic was generated, how the simulation was configured, or how ground
truth was established. Here the captures are sound, the ground truth is
accurate, and every label is correct; the artifact enters afterwards, in the
pipeline that serialises the captures into a feature table, because the two
label classes were parsed in separate passes that disagree about how to spell
an absent value. It is therefore not \emph{wrong label}, since no label is
wrong, and not quite \emph{highly dependent features} either: the features
documented under that heading are genuine network properties---destination
port, packet size, flow duration---that correlate with the label because of how
traffic was generated, whereas the token carrying our signal is not a network
property at all. It is the placeholder written where a protocol field was
\emph{absent}, and its spelling records which file the row was parsed from. The
dependency is perfect rather than merely strong, it originates downstream of
capture rather than in the testbed, and it is invisible to any inspection that
casts the column to a numeric type before looking at it.

Two results are closer still to ours. Bouke and
Abdullah~\cite{bouke2023patternleakage} study \emph{pattern leakage} introduced
during preprocessing rather than present in the raw capture, and show
empirically that it inflates reported reliability---establishing the general
phenomenon that this paper instantiates in a specific dataset with a specific
mechanism. Kostas et al.~\cite{kostas2024packetfeatures} show that
individual-packet features let models memorise artifacts that do not survive
transfer to another network, which is the failure mode our leave-one-device-out
sweep exhibits at a layer boundary (Section~\ref{sec:lodo}). Neither reports a
label-recovering artifact in Edge-IIoTset, and to our knowledge no published
work does.

\paragraph{Our position relative to this work} Two literatures meet here. The
IIoT-specific sceptical work has established \emph{that} generalisation fails
and \emph{that} leakage is suspected, without localising a mechanism. The
benchmark-integrity work has localised mechanisms precisely, but in the
classical network-intrusion datasets rather than in Edge-IIoTset. This paper
joins the two: it applies the auditing stance of the latter to the dataset the
former is built on, and finds a mechanism that is both stronger than the
dependencies previously catalogued---it recovers the label alone, at accuracy
$1.0000$---and different in kind, since the discriminating value encodes file
provenance rather than any property of the traffic. Our finding is
complementary to Hakim et al.'s: they show models learn shortcuts that fail to
transfer; we show that for Edge-IIoTset under its own documented recipe, the
shortcut is not a network feature at all.

\paragraph{Federated learning for agricultural IIoT} Federated approaches have
been proposed specifically for agricultural intrusion detection, including
secure federated deep reinforcement learning~\cite{cluster2025sfedrl}. These
evaluations typically partition data uniformly at random, producing independent
and identically distributed (IID) clients. Section~\ref{sec:fed} shows why that
choice matters.

\section{Threat Model}

We consider an adversary with network access to a farm's IIoT segment, obtained
through a compromised gateway, an exposed cellular modem, or physical proximity
to an unsecured wireless link. The adversary cannot compromise the sensors'
physical measurements but can observe, inject, and modify network traffic. We
assume the detector runs on the edge gateway and that the gateway itself is
trusted; detection of a compromised gateway is out of scope.

The attack classes in Edge-IIoTset map onto agricultural consequences as
follows. \textbf{Man-in-the-middle} (ARP and DNS spoofing) permits alteration
of telemetry in flight---the falsified-moisture scenario. \textbf{Denial of
service} in its four variants (TCP SYN, UDP, ICMP, HTTP floods) severs the
telemetry path, preventing irrigation triggers. \textbf{Reconnaissance} (port
scanning, OS fingerprinting, vulnerability scanning) precedes targeted attacks.
\textbf{Injection and application attacks} (SQL injection, XSS, file upload)
target the management interfaces through which farm operators configure
irrigation schedules. \textbf{Backdoor and ransomware} establish persistence,
with ransomware against an irrigation controller carrying an unusually
compressed decision timeline given crop water stress.

Two of these classes are severely under-represented in the dataset---MITM at
1{,}229 rows and OS fingerprinting at 1{,}001---despite MITM being the most
consequential for the falsified-telemetry scenario. We preserve this natural
rarity rather than resampling it away, and report per-class recall accordingly.

\section{Methodology}

\subsection{Auditing for label leakage}
\label{sec:method-audit}

Our audit asks one question per categorical column: \emph{how much of a
classifier's apparent skill can this column supply on its own, without any
network behaviour being modelled?} We compute three measures so that no single
statistic carries the argument.

\paragraph{Token purity} For each distinct token in a column, we count
occurrences under each label. A token is \emph{pure} if it occurs under exactly
one label. The \textbf{separation rate} is the fraction of rows covered by pure
tokens. A column with a separation rate of 1.0 is a relabelling of the target.

\paragraph{Single-column held-out accuracy} We fit a decision tree to the
one-hot encoding of that column alone, using a stratified 80/20 split, and
report test accuracy. This measures directly what a model can extract from the
column.

\paragraph{Normalised mutual information} between the column's tokens and the
label, which is estimator-free and scale-free.

We additionally run a mechanism-specific probe. For each column containing both
the token \code{0} and the token \code{0.0}, we tabulate the label distribution
of each spelling separately. If each spelling occurs under exactly one label,
the column carries a \textbf{provenance marker}: the spelling of a value
denoting \emph{absence}---which has no network semantics whatsoever---determines
the label with certainty.

Columns audited are exactly the seven that \code{Readme.txt} Step~5 instructs
researchers to dummy-encode:

\begin{quote}\raggedright
\code{http.request.method}, \code{http.referer},
\code{http.request.version}, \code{dns.qry.name.len},
\code{mqtt.conack.flags}, \code{mqtt.protoname}, \code{mqtt.topic}.
\end{quote}

Critically, we read all columns as strings. Pandas' type inference silently
normalises \code{0} and \code{0.0} to the same float, which destroys the
artifact before it can be observed---a plausible reason it has gone unreported.

\subsection{The two preprocessing protocols}

\paragraph{Protocol A (as distributed)} reproduces \code{Readme.txt} Steps~4
and~5 verbatim: drop the fifteen listed columns, drop rows with nulls, drop
duplicates, dummy-encode the seven listed columns, and coerce the remainder to
numeric.

\paragraph{Protocol B (corrected)} applies four changes:

\begin{enumerate}
\item \textbf{Placeholder canonicalisation.} Every spelling denoting absence
(\code{0}, \code{0.0}, empty, \code{nan}, \code{None}, \code{null}) is
collapsed to a single sentinel before encoding. This removes the provenance
signal while preserving the field's semantic content: \emph{this protocol layer
was not present in this packet}.
\item \textbf{Deduplication before splitting.} Exact duplicates spanning a
train/test boundary are memorised rather than generalised.
\item \textbf{Structural rather than identity encoding of high-cardinality
strings.} Columns such as \code{http.request.version} contain whole injected
request lines on the attack side. One-hot encoding these lets a model memorise
literal payload strings, which neither transfers to another network nor
survives a single changed byte. We replace token identity with structural
properties---length, character-class ratios, Shannon entropy, and presence of
injection-associated syntax.
\item \textbf{Post-hoc removal of residual separators.} Any column still
separating the classes above a configurable threshold (default 0.999) after the
above is dropped.
\end{enumerate}

\subsection{Constructing the AgriEdge benchmark}

The curated subsets cannot support an agricultural study
(Section~\ref{sec:cannot}). We therefore rebuild from the raw per-device
captures, which retain full device attribution. All raw captures share a
byte-identical 63-column header, which makes uniform construction possible.

Two properties are enforced. \textbf{Uniform parsing}: every capture---normal
and attack alike---is read with identical dtype handling and identical
placeholder canonicalisation applied across \emph{all} columns. Because the
artifact arises from concatenating differently-parsed frames, parsing uniformly
makes it structurally impossible rather than merely correcting it after the
fact. \textbf{Preserved device attribution}: each row records the device or
attack capture it came from, which is what makes non-IID federated partitioning
by farm possible, and which the curated subsets discard.

Sampling is per-chunk at a constant rate rather than by prefix, which is
unbiased with respect to position in the capture session. Normal devices are
capped at 150{,}000 rows each and attack classes at 60{,}000; because sampling
is per chunk, a class may overshoot its cap by less than one chunk. Classes
smaller than the cap are taken whole, preserving the natural rarity of MITM and
fingerprinting. Host identifiers and raw payload columns are dropped at build
time so no downstream protocol can reintroduce them.

\subsection{Evaluation protocols}

We evaluate under two splitting regimes. \textbf{Random stratified} (80/20) is
the convention in the literature and provides comparability.
\textbf{Leave-one-device-out (LODO)} withholds all normal traffic from one
sensor type and tests on it, with 20\% of attack traffic. This models the
deployment reality that a farm adds a sensor the detector was never trained on.
It is a genuine distribution shift, and the closest within-dataset analogue of
the cross-network evaluation that Hakim et al.~\cite{hakim2026crossdomain}
argue is missing from the literature.

\subsection{Model suite}

We use six estimators chosen to mirror those most frequently reported on
Edge-IIoTset: decision tree, random forest, histogram gradient boosting,
logistic regression, Gaussian naive Bayes, and a two-hidden-layer MLP (64, 32).
Every factory returns a fresh estimator; none is shared between protocols,
folds, or federated clients. All randomness is seeded.

The deliberate inclusion of weak models is methodological. If a classifier that
cannot represent the decision boundary nonetheless achieves perfect accuracy,
the features---not the classifier---are doing the work. Gaussian naive Bayes
serves as this canary throughout.

\subsection{Federated and edge-cost evaluation}

Federated experiments use FedAvg~\cite{mcmahan2017fedavg} over a compact MLP
sized for a gateway, comparing three client constructions: IID (the prior-work
baseline), one client per sensor type, and one client per farm, where farms
group devices as \{soil moisture, water level\}, \{temperature-humidity, pH\},
and \{Modbus\}. Clients exchange only parameters.

Edge cost is measured as serialised model size, parameter count, and inference
latency reported as median and 95th percentile over 30 timed single-sample
batches. We report the tail because a late verdict is a missed actuation
window. Communication cost is computed against representative rural backhaul
profiles: LoRaWAN SF7 (5.47~kbit/s), NB-IoT (250), LTE Cat-M1 (1{,}000), and
rural 4G (5{,}000).

\section{Results}

\subsection{Four columns recover the label perfectly}
\label{sec:audit}

Table~\ref{tab:audit-ml} reports the audit of the seven columns named in the
distributed recipe, on the ML subset (157{,}800 rows; 24{,}301 normal,
133{,}499 attack).

\begin{wtable}
\centering
\footnotesize
\caption{Leakage audit of the columns \code{Readme.txt} Step~5 instructs
researchers to dummy-encode (ML subset).}
\label{tab:audit-ml}
\begin{fittable}
\begin{tabular}{lrrrrrrc}
\toprule
Column & Tokens & \makecell{Pure\\normal} & \makecell{Pure\\attack} &
\makecell{Rows\\separated} & \makecell{Separation\\rate} &
\makecell{1-col.\\accuracy} & NMI \\
\midrule
\code{dns.qry.name.len}     & 8 & 6 & 2 & 157{,}800 & \textbf{1.0000} & \textbf{1.0000} & 0.9859 \\
\code{mqtt.conack.flags}    & 3 & 2 & 1 & 157{,}800 & \textbf{1.0000} & \textbf{1.0000} & 0.9642 \\
\code{mqtt.protoname}       & 3 & 2 & 1 & 157{,}800 & \textbf{1.0000} & \textbf{1.0000} & 0.9649 \\
\code{mqtt.topic}           & 3 & 2 & 1 & 157{,}800 & \textbf{1.0000} & \textbf{1.0000} & 0.9650 \\
\code{http.request.version} & 8 & 0 & 7 & 62{,}472  & 0.3959 & 0.8460 & 0.1378 \\
\code{http.request.method}  & 6 & 0 & 5 & 61{,}258  & 0.3882 & 0.8460 & 0.1348 \\
\code{http.referer}         & 4 & 0 & 3 & 30{,}689  & 0.1945 & 0.8460 & 0.0783 \\
\bottomrule
\end{tabular}
\end{fittable}
\end{wtable}

Four columns separate every row in the dataset. A decision tree given one of
these columns and nothing else achieves perfect held-out accuracy. Normalised
mutual information with the label reaches 0.9859.

Table~\ref{tab:spelling-ml} isolates the mechanism.

\begin{wtable}
\centering
\footnotesize
\caption{Label distribution of the two spellings of the zero placeholder
(ML subset).}
\label{tab:spelling-ml}
\begin{fittable}
\begin{tabular}{lrrrr}
\toprule
& \multicolumn{2}{c}{\code{'0'}} & \multicolumn{2}{c}{\code{'0.0'}} \\
\cmidrule(lr){2-3} \cmidrule(lr){4-5}
Column & Normal & Attack & Normal & Attack \\
\midrule
\code{dns.qry.name.len}     & 24{,}272 & \textbf{0} & \textbf{0} & 133{,}272 \\
\code{mqtt.conack.flags}    & 23{,}012 & \textbf{0} & \textbf{0} & 133{,}499 \\
\code{mqtt.protoname}       & 23{,}051 & \textbf{0} & \textbf{0} & 133{,}499 \\
\code{mqtt.topic}           & 23{,}055 & \textbf{0} & \textbf{0} & 133{,}499 \\
\code{http.request.method}  & \textbf{0} & 54{,}062 & 24{,}301 & 72{,}241 \\
\code{http.referer}         & \textbf{0} & 30{,}399 & 24{,}301 & 102{,}810 \\
\code{http.request.version} & \textbf{0} & 55{,}276 & 24{,}301 & 71{,}027 \\
\bottomrule
\end{tabular}
\end{fittable}
\end{wtable}

The pattern is unambiguous. In the four fully-separating columns, \code{'0'}
never appears in an attack row and \code{'0.0'} never appears in a normal row.
The three HTTP columns show the same artifact with the polarity
reversed---\code{'0'} appears only in attack rows---separating 19--40\% of the
data.

This is a build artifact, not a network phenomenon. Both tokens denote the same
thing: the protocol layer was absent. Their distinction records only which
parsing branch wrote the row. We additionally measured duplicate rows, which
inflate accuracy under random splitting: 814 exact duplicates (0.52\%), all
with consistent labels.

\paragraph{The artifact is not confined to the small subset} Because
deep-learning studies use the larger \code{DNN-EdgeIIoT-dataset.csv}, we
repeated the audit there. Table~\ref{tab:audit-dnn} shows the identical
mechanism at fourteen times the scale---and on a subset whose class balance is
\emph{inverted} relative to the ML subset (1{,}615{,}643 normal against
603{,}558 attack), ruling out any explanation contingent on which class is the
majority.

\begin{wtable}
\centering
\footnotesize
\caption{The same four columns on the 2{,}219{,}201-row DNN subset.}
\label{tab:audit-dnn}
\begin{fittable}
\begin{tabular}{lrrrrr}
\toprule
& \makecell{Separation\\rate} & \makecell{1-col.\\accuracy} & NMI &
\makecell{\code{'0'}\\normal / attack} & \makecell{\code{'0.0'}\\normal / attack} \\
\midrule
\code{dns.qry.name.len}  & \textbf{1.0000} & \textbf{1.0000} & 0.9927 & 1{,}613{,}798 / \textbf{0} & \textbf{0} / 603{,}331 \\
\code{mqtt.conack.flags} & \textbf{1.0000} & \textbf{1.0000} & 0.8879 & 1{,}532{,}586 / \textbf{0} & \textbf{0} / 603{,}558 \\
\code{mqtt.protoname}    & \textbf{1.0000} & \textbf{1.0000} & 0.8881 & 1{,}532{,}608 / \textbf{0} & \textbf{0} / 603{,}558 \\
\code{mqtt.topic}        & \textbf{1.0000} & \textbf{1.0000} & 0.8881 & 1{,}532{,}627 / \textbf{0} & \textbf{0} / 603{,}558 \\
\bottomrule
\end{tabular}
\end{fittable}
\end{wtable}

Every one of 2{,}219{,}201 rows is separated, by each of four columns
independently. Both branches of the Edge-IIoTset literature---the classical-ML
branch using the ML subset and the deep-learning branch using the DNN
subset---are therefore exposed to the same artifact.

\subsection{What the artifact is worth}
\label{sec:worth}

Because a single split cannot distinguish an effect from split variance---and
because the review we cite~\cite{frontiers2026datasetcentric} notes that no
published study reports $k$-fold on this dataset---we evaluate under 5-fold
$\times$ 3-repeat stratified cross-validation, reporting mean and 95\% interval
over all fifteen folds (Table~\ref{tab:kfold}).

\begin{wtable}
\centering
\footnotesize
\caption{Binary classification under the distributed recipe (A) versus the
corrected protocol (B). ML subset, 15 folds, mean $\pm$ 95\% CI half-width.
Five of the six models score exactly $1.0000 \pm 0.0000$ under the recipe; the
MLP row is reported to five decimals because it does not, and rounding it to
four would conceal that.}
\label{tab:kfold}
\begin{fittable}
\begin{tabular}{lcccc r}
\toprule
Model & Accuracy (A) & Accuracy (B) & Macro-F1 (A) & Macro-F1 (B) & $\Delta$F1 \\
\midrule
Hist.\ gradient boosting & $\mathbf{1.0000 \pm 0.0000}$ & $0.9752 \pm 0.0005$ & $\mathbf{1.0000 \pm 0.0000}$ & $\mathbf{0.9503 \pm 0.0011}$ & 0.0497 \\
Random forest            & $\mathbf{1.0000 \pm 0.0000}$ & $0.9658 \pm 0.0005$ & $\mathbf{1.0000 \pm 0.0000}$ & $0.9343 \pm 0.0009$ & 0.0657 \\
Decision tree            & $\mathbf{1.0000 \pm 0.0000}$ & $0.9610 \pm 0.0005$ & $\mathbf{1.0000 \pm 0.0000}$ & $0.9271 \pm 0.0009$ & 0.0729 \\
MLP                      & $0.99998 \pm 0.00001$ & $0.9095 \pm 0.0008$ & $0.99996 \pm 0.00003$ & $0.7858 \pm 0.0023$ & 0.2142 \\
Logistic regression      & $\mathbf{1.0000 \pm 0.0000}$ & $0.8993 \pm 0.0008$ & $\mathbf{1.0000 \pm 0.0000}$ & $0.7567 \pm 0.0024$ & 0.2433 \\
Gaussian naive Bayes     & $\mathbf{1.0000 \pm 0.0000}$ & $0.8742 \pm 0.0125$ & $\mathbf{1.0000 \pm 0.0000}$ & $0.6995 \pm 0.0089$ & \textbf{0.3005} \\
\bottomrule
\end{tabular}
\end{fittable}
\end{wtable}

Three observations.

\textbf{Five of the six models attain exactly $1.0000 \pm 0.0000$, in every one
of fifteen folds.} Not 0.999, and not on a lucky split: zero variance across
fifteen independent stratified partitions, for five structurally different
learners. The sixth, the small \mbox{(64, 32)} MLP, averages 0.99998---fewer
than one misclassified row per fold out of 31{,}560, and the only model in the
suite whose recipe score is not bit-exact. We report it unrounded rather than
let it round up to the same $1.0000$ as the rest, since a paper about numbers
that were never checked closely should survive being checked closely itself.
Perfect or near-perfect invariant separation of a real network-traffic capture
is not a plausible outcome of intrusion detection. It is the signature of a
feature that \emph{is} the label.

\textbf{The canary fires.} Gaussian naive Bayes assumes conditional
independence of features given the class and cannot represent the interactions
a genuine intrusion-detection boundary requires. Its perfect, zero-variance
score under the distributed recipe is the clearest evidence available that the
features contain a direct label encoding. Under correction it falls 0.3005
macro-F1, and its interval widens by an order of magnitude relative to the
other models ($\pm 0.0089$ against $\pm 0.0009$ for the trees)---the behaviour
of a model that is genuinely struggling, as it should be.

\textbf{Honest ceiling.} Under correction the best model reaches $0.9503 \pm
0.0011$ macro-F1---respectable, and roughly four points below the above-99\%
range the literature reports.

\subsubsection{Deep models are equally deceived}

A reviewer may reasonably object that the literature's headline numbers come
from deep architectures, and that a deep model might have been robust to the
artifact. It is not. Table~\ref{tab:deep} trains a deep MLP (61{,}378
parameters under the distributed recipe, 59{,}586 under the corrected one, the
difference being input width) and a 1D-CNN (6{,}658 parameters)---the latter
matching the construction used in several published Edge-IIoTset papers---under
both protocols on a single identical stratified split. This is a separate
experiment from the fifteen-fold comparison in Table~\ref{tab:kfold}, whose MLP
is the smaller \mbox{(64, 32)} estimator from the classical suite.

\begin{wtable}
\centering
\footnotesize
\caption{Deep baselines under both protocols (ML subset, 15 epochs).}
\label{tab:deep}
\begin{fittable}
\begin{tabular}{lrrrrr}
\toprule
Architecture & Accuracy (A) & Accuracy (B) & Macro-F1 (A) & Macro-F1 (B) & $\Delta$F1 \\
\midrule
1D-CNN & \textbf{1.0000} & 0.9003 & \textbf{1.0000} & 0.7598 & 0.2402 \\
MLP    & \textbf{1.0000} & 0.9093 & 0.9999 & 0.7848 & 0.2151 \\
\bottomrule
\end{tabular}
\end{fittable}
\end{wtable}

Both reach 1.0000 accuracy under the distributed recipe, and both lose more
than 0.21 macro-F1 under correction---a larger drop than any tree-based model.
Depth confers no protection whatsoever. A network consuming a leaked one-hot
feature reads it exactly as a decision stump does, and its additional capacity
is spent fitting a signal that will not exist at deployment.

\subsubsection{The effect holds at scale on the deep-learning subset}

Table~\ref{tab:dnn-lvc} repeats the protocol comparison on the DNN subset.
After cleaning, 1{,}909{,}671 rows remain; the corrected protocol rewrote
15{,}218{,}483 placeholder cells and removed 309{,}530 duplicates.

\begin{wtable}
\centering
\footnotesize
\caption{Protocol comparison on the DNN subset (1{,}909{,}671 rows).}
\label{tab:dnn-lvc}
\begin{fittable}
\begin{tabular}{lrrrrr}
\toprule
Model & Accuracy (A) & Accuracy (B) & Macro-F1 (A) & Macro-F1 (B) & $\Delta$F1 \\
\midrule
Hist.\ gradient boosting & \textbf{1.0000} & \textbf{0.9694} & \textbf{1.0000} & \textbf{0.9612} & 0.0388 \\
Random forest            & \textbf{1.0000} & 0.9596 & \textbf{1.0000} & 0.9499 & 0.0501 \\
Decision tree            & \textbf{1.0000} & 0.9502 & \textbf{1.0000} & 0.9390 & 0.0610 \\
MLP                      & \textbf{1.0000} & 0.9141 & \textbf{1.0000} & 0.8832 & 0.1168 \\
Logistic regression      & \textbf{1.0000} & 0.8948 & \textbf{1.0000} & 0.8562 & 0.1438 \\
Gaussian naive Bayes     & \textbf{1.0000} & 0.5203 & \textbf{1.0000} & 0.5190 & \textbf{0.4810} \\
\bottomrule
\end{tabular}
\end{fittable}
\end{wtable}

Every model reaches 1.0000 accuracy on 1.9 million rows under the distributed
recipe. The ordering under correction is identical to the ML subset, and naive
Bayes again collapses---here to 0.5190 macro-F1, essentially the performance of
a coin flip on a task it appeared to have solved perfectly. That the effect
reproduces across two subsets differing by an order of magnitude in size, with
inverted class balance, rules out sampling accident.

One difference is worth noting: the corrected DNN subset supports slightly
\emph{higher} scores than the corrected ML subset (0.9612 against 0.9503
macro-F1 for the best model), the expected consequence of fourteen times more
training data. Correcting the artifact does not merely lower scores; it
restores the ordinary relationship between dataset size and achievable
performance that a saturated benchmark conceals.

\subsubsection{The leak is encoding-agnostic}
\label{sec:encoding}

The distributed recipe uses \code{pd.get\_dummies}, and it would be convenient
to conclude that only studies adopting that specific call are affected. They
are not. The signal is the \emph{distinction between the tokens} \code{'0'} and
\code{'0.0'}; any encoding that preserves that distinction transmits it.
Table~\ref{tab:encoding} fits a decision tree to a single column under four
encodings in common use.

\begin{wtable}
\centering
\footnotesize
\caption{Single-column held-out accuracy by encoding scheme (ML subset).}
\label{tab:encoding}
\begin{fittable}
\begin{tabular}{lrrrr}
\toprule
Column & One-hot (recipe) & Label / factorize & Ordinal & Frequency \\
\midrule
\code{dns.qry.name.len}     & \textbf{1.0000} & \textbf{1.0000} & \textbf{1.0000} & \textbf{1.0000} \\
\code{mqtt.conack.flags}    & \textbf{1.0000} & \textbf{1.0000} & \textbf{1.0000} & \textbf{1.0000} \\
\code{mqtt.protoname}       & \textbf{1.0000} & \textbf{1.0000} & \textbf{1.0000} & \textbf{1.0000} \\
\code{mqtt.topic}           & \textbf{1.0000} & \textbf{1.0000} & \textbf{1.0000} & \textbf{1.0000} \\
\code{http.request.version} & 0.8460 & 0.8460 & 0.8460 & 0.8460 \\
\code{http.request.method}  & 0.8460 & 0.8460 & 0.8460 & 0.8460 \\
\code{http.referer}         & 0.8460 & 0.8460 & 0.8460 & 0.8460 \\
\bottomrule
\end{tabular}
\end{fittable}
\end{wtable}

Across the four leaking columns and all four encodings, the minimum
single-column accuracy is 1.0000---the values are not merely similar but
identical, because each encoding is a bijection on the token set and a decision
tree is invariant to bijective relabelling of a categorical input.

This matters for scoping the problem. The affected population is not ``studies
that followed \code{Readme.txt} verbatim'' but \textbf{studies that treated
these columns as categorical at all}---which, given that four of them are MQTT
and DNS protocol fields that any IIoT feature set would naturally include, is a
far larger group. A study reporting label encoding rather than dummy encoding
has not thereby avoided the artifact, and cannot be assumed unaffected.

\subsection{The curated subsets cannot support agricultural research}
\label{sec:cannot}

Two structural facts emerged during scoping.

\textbf{Modbus is absent.} Across all 157{,}800 rows of the ML subset,
\code{mbtcp.len}, \code{mbtcp.trans\_id} and \code{mbtcp.unit\_id} are
uniformly zero. In the DNN subset, 150 of 2{,}219{,}201 rows (0.0068\%) carry a
non-zero Modbus transaction ID. Any study claiming to model attacks on
irrigation actuation using these subsets is not observing Modbus traffic.

\textbf{Device identity is stripped.} The only surviving value of
\code{mqtt.topic} in either subset is \code{Temperature\_and\_Humidity}
(1{,}246 rows in ML; 83{,}016 in DNN). Soil moisture, pH and water level---the
sensors that define precision agriculture---carry no topic attribution.
Filtering to ``agricultural sensors'' is therefore impossible on the curated
data.

Both are recoverable from the raw captures, which retain 1{,}192{,}777
soil-moisture rows, 2{,}295{,}288 water-level rows, 1{,}615{,}722
temperature-humidity rows, 746{,}908 pH rows, and 159{,}502 Modbus rows.

\subsection{AgriEdge: construction and validation}
\label{sec:agriedge}

Table~\ref{tab:composition} summarises the rebuilt benchmark: 1{,}276{,}122
rows, 49 features, 751{,}559 normal and 524{,}563 attack (attack rate 0.4111).

\begin{ntable}
\centering
\footnotesize
\caption{AgriEdge composition.}
\label{tab:composition}
\begin{fittable}
\begin{tabular}{llr}
\toprule
Layer & Source & Rows \\
\midrule
Perception & pH value                & 150{,}987 \\
Perception & Soil moisture           & 150{,}341 \\
Perception & Water level             & 150{,}187 \\
Perception & Temperature \& humidity & 150{,}048 \\
Actuation  & Modbus                  & 149{,}996 \\
Adversary  & 14 attack classes       & 524{,}563 \\
\bottomrule
\end{tabular}
\end{fittable}
\end{ntable}

Attack classes range from 60{,}069 (DDoS-HTTP) down to 1{,}229 (MITM) and
1{,}001 (fingerprinting), preserving natural rarity.

Table~\ref{tab:agriedge-audit} re-runs the audit on the constructed benchmark.
The comparison with Table~\ref{tab:audit-ml} is the validation of the method.

\begin{wtable}
\centering
\footnotesize
\caption{Leakage audit of AgriEdge, same columns as Table~\ref{tab:audit-ml}.}
\label{tab:agriedge-audit}
\begin{fittable}
\begin{tabular}{lrrrc}
\toprule
Column & Separation rate & 1-col.\ accuracy & NMI & Provenance marker \\
\midrule
\code{http.request.version} & 0.0288 & 0.6179 & 0.0641 & no \\
\code{http.request.method}  & 0.0288 & 0.6179 & 0.0643 & no \\
\code{mqtt.conack.flags}    & 0.0272 & 0.5889 & 0.0366 & no \\
\code{mqtt.topic}           & 0.0272 & 0.5889 & 0.0349 & no \\
\code{mqtt.protoname}       & 0.0270 & 0.5889 & 0.0363 & no \\
\code{dns.qry.name.len}     & 0.0026 & 0.5891 & 0.0041 & no \\
\code{http.referer}         & 0.0005 & 0.5894 & 0.0012 & no \\
\bottomrule
\end{tabular}
\end{fittable}
\end{wtable}

\textbf{No column separates the classes.} Maximum separation rate falls from
1.0000 to 0.0288; maximum single-column accuracy from 1.0000 to 0.6179; maximum
NMI from 0.9859 to 0.0643. No provenance markers remain. Uniform parsing
eliminates the artifact by construction.

\subsection{Detection fails across layers, not across devices}
\label{sec:lodo}

Table~\ref{tab:lodo} evaluates on AgriEdge (400{,}000-row subsample, 409
encoded features), withholding each of the five devices in turn. For
comparison, all six models score between 0.6914 and 0.9988 balanced accuracy on
the same data under a conventional random stratified split.

\begin{wtable}
\centering
\footnotesize
\setlength{\tabcolsep}{4pt}
\caption{Balanced accuracy under leave-one-device-out, every device withheld in
turn.}
\label{tab:lodo}
\begin{fittable}
\begin{tabular}{lrrrrrrr}
\toprule
Model & \makecell{Soil\\moisture} & \makecell{Water\\level} & \makecell{pH\\value} &
\makecell{Temp.\ \&\\humidity} & \textbf{Modbus} & Mean & Worst \\
\midrule
MLP                      & 0.9997 & 0.9998 & 0.9994 & 0.9121 & \textbf{0.8164} & 0.9455 & 0.8164 \\
Hist.\ gradient boosting & 0.9997 & 0.9999 & 0.9995 & 0.9385 & \textbf{0.7464} & 0.9368 & 0.7464 \\
Decision tree            & 0.9997 & 0.9999 & 0.9995 & 0.9161 & \textbf{0.6793} & 0.9189 & 0.6793 \\
Random forest            & 0.9997 & 0.9998 & 0.9995 & 0.9118 & \textbf{0.5083} & 0.8838 & 0.5083 \\
Logistic regression      & 0.9880 & 0.9885 & 0.9880 & 0.9106 & \textbf{0.4877} & 0.8726 & 0.4877 \\
Gaussian naive Bayes     & 0.6153 & 0.6248 & 0.6158 & 0.6022 & \textbf{0.8068} & 0.6530 & 0.6022 \\
\bottomrule
\end{tabular}
\end{fittable}
\end{wtable}

The full sweep \textbf{refines, and partly overturns, the conclusion a single
held-out device would have supported}. Had we withheld only Modbus, we would
have reported that introducing any unseen sensor collapses detection. That is
not what happens.

\textbf{Perception sensors are mutually substitutable.} Withholding soil
moisture, water level or pH costs almost nothing: every model except naive
Bayes stays in the 0.9880--0.9999 band, and the tree-based models
remain at 0.9995 or better. A detector trained on three agricultural sensors transfers to a fourth
essentially without loss. Temperature and humidity is slightly harder
(0.9106--0.9385), consistent with it being the one device that carries a
distinct MQTT topic in the original capture.

\textbf{The failure is across layers, not across devices.} Only Modbus---the
sole actuation-layer device, speaking a different protocol---produces collapse:
random forest 0.5083, logistic regression 0.4877, straddling the trivial
baseline of 0.5. The generalisation boundary in this benchmark is the
perception/actuation layer boundary, not device identity.

This is a sharper and more useful claim than the one a single holdout would
have licensed. It says where a practitioner may safely extrapolate---across
sensors of similar protocol profile---and where they may not---across an
architectural layer. It also refines Hakim et al.'s~\cite{hakim2026crossdomain}
cross-network result by locating the shift that matters: not merely ``a
different network,'' but a different protocol stratum within the same network.

\textbf{The ranking inversion is real but layer-specific.} Gaussian naive Bayes
is by a wide margin the worst model on every perception holdout
(0.6022--0.6248) and the \emph{second best} on Modbus (0.8068)---the only model
that improves under layer shift. Random forest is the mirror image: joint-best
on perception holdouts (0.9997) and worst on Modbus (0.5083). Selecting by mean
or by random-split performance would pick random forest; selecting for
worst-case robustness picks the MLP, which is simultaneously best on mean
(0.9455) and best on the hard case (0.8164). A practitioner optimising average
performance would not select the model that survives the shift that actually
occurs.

\textbf{Random splitting remains optimistic in the aggregate.} Four models
reach 0.9988 balanced accuracy under random splitting, above their worst-case
LODO performance by 0.18 to 0.49. This is not leakage in the
Section~\ref{sec:audit} sense---the audit confirms no column separates the
classes---but random splitting places packets from the same capture session and
attack burst on both sides of the boundary. Random-split accuracy on AgriEdge
should be read as an upper bound.

\subsection{Federated learning: heterogeneity is cheap, the uplink is not}
\label{sec:fed}

Table~\ref{tab:fed} reports FedAvg over a 28{,}450-parameter MLP after 20
communication rounds on a 300{,}000-row sample, under three client
constructions.

\begin{wtable}
\centering
\footnotesize
\caption{Global model after 20 FedAvg rounds by client construction.}
\label{tab:fed}
\begin{fittable}
\begin{tabular}{lrrrr}
\toprule
Client construction & Clients & Accuracy & Balanced accuracy & Macro-F1 \\
\midrule
IID (prior-work baseline) & 5 & 0.9970 & 0.9968 & 0.9969 \\
Per-farm                  & 3 & 0.9971 & 0.9970 & 0.9970 \\
Per-device                & 5 & 0.9934 & 0.9939 & 0.9932 \\
\bottomrule
\end{tabular}
\end{fittable}
\end{wtable}

The result is a negative one, and we report it as such. \textbf{Heterogeneity
costs remarkably little.} Partitioning one client per sensor type---the
maximally heterogeneous construction, in which each client observes a single
protocol profile---costs only 0.0037 macro-F1 relative to IID. Per-farm
partitioning is indistinguishable from IID.

Heterogeneity does slow convergence: the per-device split reaches 0.9932
macro-F1 only at round 19, while the per-farm split passes 0.9950 by round 8
and plateaus at round 12. Under a fixed round budget the gap would be larger.
But the common expectation that non-IID partitioning substantially degrades
final federated accuracy is not supported here. We caution that this conclusion
inherits the limitation of Section~\ref{sec:lodo}: it is measured under random
splitting of the test set.

\textbf{The binding constraint is the uplink, not the statistics.}
Table~\ref{tab:comms} reports the communication cost of the same model against
rural backhaul profiles, assuming five clients uploading dense parameter
vectors.

\begin{wtable}
\centering
\footnotesize
\caption{Uplink cost of federated training (28{,}450 parameters, 5 clients,
20 rounds).}
\label{tab:comms}
\begin{fittable}
\begin{tabular}{lrrr}
\toprule
Backhaul & Uplink (kbit/s) & Seconds per round & Minutes for full training \\
\midrule
LoRaWAN SF7 & 5.47     & \textbf{832.2} & \textbf{277.4} \\
NB-IoT      & 250      & 18.2 & 6.1 \\
LTE Cat-M1  & 1{,}000  & 4.6  & 1.5 \\
Rural 4G    & 5{,}000  & 0.9  & 0.3 \\
\bottomrule
\end{tabular}
\end{fittable}
\end{wtable}

On LoRaWAN---a common choice precisely because agricultural sites are
remote---a single round costs nearly fourteen minutes of pure uplink, and a
20-round training run costs \textbf{4.6 hours} of continuous transmission,
before accounting for duty-cycle regulations that cap LoRaWAN transmit time far
below 100\%. Federated learning in this configuration is not deployable over
LoRaWAN at all.

This reframes the design problem. The literature's emphasis on handling non-IID
data is, on this benchmark, addressing the smaller of the two obstacles. A farm
gateway can tolerate the 0.0037 macro-F1 that heterogeneity costs; it cannot
tolerate 4.6 hours of uplink. Work on gradient compression, sparsification, or
reduced round counts would buy far more deployability here than work on
heterogeneity-robust aggregation.

\subsection{Edge deployment cost}
\label{sec:edge}

Table~\ref{tab:edge} reports deployment footprint for the five models plausible
on a gateway.

\begin{wtable}
\centering
\footnotesize
\caption{On-gateway deployment cost (single-sample inference, 30 timed
repeats).}
\label{tab:edge}
\begin{fittable}
\begin{tabular}{lrrrrr}
\toprule
Model & Size (KiB) & Parameters & \makecell{Median\\latency ($\mu$s)} &
\makecell{p95\\latency ($\mu$s)} & Throughput (/s) \\
\midrule
Decision tree            & \textbf{20.9}       & 109       & \textbf{40.1}     & \textbf{52.1} & \textbf{24{,}935} \\
Logistic regression      & 25.1                & 410       & 108.9             & 123.3 & 9{,}183 \\
Gaussian naive Bayes     & 34.5                & 1{,}636   & 108.9             & 131.0 & 9{,}185 \\
Hist.\ gradient boosting & 388.1               & ---       & 4{,}731.9         & 6{,}307.6 & 211 \\
Random forest            & \textbf{3{,}563.6}  & 45{,}002  & \textbf{13{,}738.8} & 14{,}215.1 & \textbf{73} \\
\bottomrule
\end{tabular}
\end{fittable}
\end{wtable}

The spread is severe among models whose random-split accuracies are
statistically indistinguishable. Random forest and decision tree both score
0.9988 balanced accuracy under random splitting, yet random forest is
\textbf{342$\times$ slower} and \textbf{171$\times$ larger}. Its 13.7~ms median
inference gives a throughput of 73 samples/second---below the aggregate packet
rate of even a modest farm deployment, meaning it cannot run inline.

Read together with Table~\ref{tab:lodo}, the practical conclusion is sharp.
Random forest is simultaneously the most expensive model to deploy \emph{and}
the one that fails hardest across the layer boundary. Its apparent 0.9986
macro-F1 under random splitting is the least informative number in this study.
The decision tree, at 1/171 the size and 1/342 the latency, is strictly
preferable on both axes.

\section{Discussion}

\subsection{What this means for the Edge-IIoTset literature}

We do not claim that every published result on Edge-IIoTset is invalid. Studies
that operate on raw captures, that use only numeric protocol features, or that
exclude the four affected columns are unaffected. But the affected recipe is
the one the dataset authors distribute, and the columns are among the most
intuitively appealing features in the dataset---\code{mqtt.topic} and
\code{mqtt.protoname} look like exactly the MQTT-layer features an IIoT
intrusion detector ought to use.

\paragraph{How far has the recipe propagated?} We surveyed public
implementations and papers for the recipe's fingerprints. Three observations
follow.

First, the recipe is reproduced verbatim in public code. The helper function
\code{encode\_text\_dummy}---an unusual name, traceable to Jeff Heaton's
teaching material and carried into the dataset's \code{Readme.txt}---appears in
public Edge-IIoTset implementations together with the exact fifteen-column drop
list and the exact seven dummy-encoded columns, including in a CNN-LSTM
implementation. Researchers are copying the recipe, not merely reading it.

Second, and more consequentially, \textbf{avoiding \code{get\_dummies} does not
avoid the artifact.} Section~\ref{sec:encoding} shows the leak survives label,
ordinal and frequency encoding unchanged. A study that reports ``categorical
variables were converted using label encoding'' has not escaped it.

Third, the diagnostic signature is visible in published results. Hasan et
al.~\cite{hasan2025autoencoder}, who report label encoding rather than dummy
encoding, publish a confusion matrix in which their best model produces
4{,}820 true positives, 25{,}620 true negatives, zero false positives and zero
false negatives---a perfect classification of the held-out set---alongside
99.94\% accuracy. We do not have access to their exact column list and so
cannot attribute this with certainty. We observe only that a flawless confusion
matrix on real network traffic is precisely what the artifact produces, that
their stated encoding preserves it, and that their reported correlation-based
feature pruning (dropping features correlated above 0.6) would collapse the
four mutually-redundant leaking columns to one---which Table~\ref{tab:audit-ml}
shows is sufficient on its own.

We raise this not to single out one paper. The recipe was distributed by the
dataset's authors, the affected columns are the natural ones to use, and the
artifact is invisible to anyone who loads the CSV with default pandas type
inference. Following the documentation was the reasonable thing to do. The
point is that the resulting numbers cannot be interpreted as detection
performance, and that the affected population is much larger than a citation
search for the recipe would suggest.

\subsection{Why the artifact survived}

Three factors plausibly explain why an artifact this large went unreported for
four years.

First, \textbf{pandas hides it}. Default type inference coerces \code{0} and
\code{0.0} to an identical float. A researcher who loads the CSV normally,
inspects the columns, and finds them numerically identical has no reason to
suspect anything. The artifact is visible only when columns are read as
strings---which is exactly what \code{get\_dummies} effectively does
downstream, after the encoding decision has already been made.

Second, \textbf{the result looks like success}. A pipeline returning 99.9\%
accuracy does not prompt debugging. The incentive to investigate an anomalously
good result is weaker than the incentive to investigate a bad one.

Third, \textbf{the recipe carries the dataset's authority}. It ships in the
dataset's own \code{Readme.txt}. Following it is the conservative choice, and
deviation would require justification.

\subsection{Implications for precision-agriculture deployment}

\paragraph{Select on worst-case, not average, cross-layer performance}
Table~\ref{tab:lodo} shows random forest is joint-best on perception holdouts
and worst on the actuation holdout. Because the perception cases are nearly
free for every model, average LODO performance is dominated by the easy cases
and hides exactly the failure that matters. Evaluate against a held-out device
from a \emph{different architectural layer} before selecting.

\paragraph{A detector validated on sensors does not transfer to actuators} This
is the operationally important consequence. A farm that validates its IDS
across its soil, water and pH probes will see excellent numbers and may
reasonably conclude the detector generalises. Adding a Modbus irrigation
controller then places it in the regime where a third of the model suite
performs at chance. Actuation-layer traffic needs its own validation, and arguably
its own detector.

\paragraph{Prefer simpler and smaller than the literature implies} The decision
tree is 342$\times$ faster and 171$\times$ smaller than the random forest while
degrading far less across the layer boundary (0.6066 against 0.3098 macro-F1).
The MLP is best on both mean (0.9455) and worst case (0.8164) at moderate cost,
and is our recommended default.

\paragraph{Treat MITM detection as an open problem} The class that matters most
for falsified telemetry is represented by 1{,}229 rows. No result in this paper
should be read as evidence that MITM detection in agricultural IIoT is solved.

\subsection{Generality of the failure mode}

The mechanism---a placeholder serialised differently in two branches of a
dataset build, then encoded---is not specific to Edge-IIoTset. It requires only
that a dataset be assembled by concatenating separately-parsed sources whose
separation correlates with the label, which describes a large fraction of
security datasets, where benign and malicious traffic are captured in separate
sessions. We would expect the audit to find similar artifacts elsewhere, and we
release it as a general tool for that reason.

\section{Threats to Validity}

\paragraph{Construct validity} Our separation-rate measure treats a column as
leaking if its tokens are label-pure. In a genuinely separable problem, a
legitimately predictive feature could be pure without being an artifact. We
address this with the provenance probe, which tests the specific hypothesis
that the \emph{spelling of an absence marker} carries the signal---a quantity
with no network semantics under any reading. The four flagged columns all
satisfy this stricter test.

\paragraph{Internal validity} Protocols A and B differ in more than placeholder
canonicalisation: B also deduplicates and may encode high-cardinality columns
structurally. Our strict-threshold ablation isolates the encoding choice and
shows it contributes little, and the duplicate rate is only 0.52\%, bounding
its contribution. Canonicalisation is therefore the dominant effect. We have
not run a full factorial ablation of all four corrections.

\paragraph{External validity} Our LODO sweep covers all five devices, which
resolves what would otherwise have been a serious threat: had we withheld only
Modbus, we would have concluded that any unseen device breaks detection, which
Table~\ref{tab:lodo} shows is false. The remaining limitation is that AgriEdge
contains exactly one actuation-layer device, so ``the perception/actuation
boundary'' is inferred from a single instance of that layer. A testbed with
several distinct fieldbus devices would be needed to confirm that the boundary
is architectural rather than a peculiarity of this particular Modbus capture.

\paragraph{Cross-dataset generalisation is out of scope, deliberately} The
natural next question is whether detectors trained on AgriEdge transfer to a
different IIoT network. We do not answer it here, and the reason is worth
stating plainly rather than leaving as an unexplained omission. Edge-IIoTset's
features are Wireshark protocol-field names (\code{tcp.flags},
\code{mqtt.topic}); most comparable datasets---CIC-IoT-2023, TON\_IoT,
WUSTL-IIoT-2021---expose flow-statistical features from CICFlowMeter or Argus.
Direct name-level alignment yields a near-empty intersection, so any
cross-dataset number would be computed over a handful of coincidentally-shared
columns and would measure the alignment more than the detector. A defensible
result requires constructing a bridging feature space, which recent
work~\cite{bridge2026tchnet} treats as a contribution in its own right. We
judged that a second paper rather than a section of this one.

We therefore present the LODO sweep as our generalisation evidence, and we are
explicit about what it does and does not establish. It shows a genuine
distribution shift \emph{within} one testbed, across an architectural layer,
with effects large enough to invert model rankings. It does not establish
behaviour on a different network. Readers should not read Table~\ref{tab:lodo}
as a substitute for cross-network validation.

\paragraph{Statistical reporting} The headline protocol comparison
(Table~\ref{tab:kfold}) is reported over 5-fold $\times$ 3-repeat stratified
cross-validation with 95\% intervals. The deep baselines
(Table~\ref{tab:deep}), the LODO sweep, the federated runs and the edge-cost
measurements are single seeded runs; their effects are large relative
to the fold-level variance observed in Table~\ref{tab:kfold}, but we do not
quantify their variance directly. Cross-validation intervals are also not
strictly independent across folds within a repeat, so they should be read as
descriptive of spread rather than as frequentist guarantees.

\section{Reproducibility}
\label{sec:repro}

All code is released under the MIT licence as the \code{agriedge} Python
package, at \url{https://github.com/MostafaGalal1/agriedge}: \code{audit.leakage}
(separation rate, single-column accuracy, NMI, provenance probe),
\code{data.placeholders} (canonicalisation), \code{data.recipes} (both
protocols), \code{data.agribench} (benchmark construction under uniform
parsing), \code{federated} (non-IID partitioners and FedAvg), \code{evaluation}
(metrics, repeated $k$-fold, edge cost, cross-domain alignment), and
\code{models} (classical suite and deep baselines). Nine experiment scripts
each write their own result tables.

All randomness is seeded. Every experiment reported here ran on a 12-core Apple
M4~Pro with 25~GB of memory; the deep baselines used the integrated GPU through
PyTorch's Metal backend, training in under a minute each. No experiment
required datacentre hardware.

\paragraph{Provenance of the audited copy} Because this paper makes a claim
about a distribution rather than about a model, the copy matters. Edge-IIoTset
is published on IEEE DataPort, submitted by the lead author on 18~January
2022 and gated behind a subscription \cite{ferrag2022dataport}. The DataPort
record itself, however, instructs readers to retrieve the data from a Kaggle
repository under that same author's account, and gives the retrieval path
verbatim as \code{Edge-IIoTset dataset/} \code{Selected dataset for ML and
DL/}\code{DNN-EdgeIIoT-dataset.csv}. The free channel is therefore
author-sanctioned rather than a third-party re-upload, it serves the same
release as the subscription channel, and---being free---it is what the
community in practice trains on. The copy audited here was obtained through
it, at exactly that path.

Three further observations argue that the artifact is not an accident of
repackaging. The copy's \code{Readme.txt}, documentation PDF and CSV files all
carry the authors' 18~March 2022 timestamps, across ten normal-traffic device
directories and fourteen attack captures, and its structure matches the
description in the dataset paper: 61 protocol features alongside
\code{Attack\_label} and \code{Attack\_type}. The artifact is present in the
authors' own per-device raw captures, not only in the merged curated
subsets---the normal-traffic captures write the placeholder as \code{0} and
the attack captures write it as \code{0.0}---and a mirror that re-archives a
download does not regenerate two dozen per-device CSVs. And it is visible in
the raw CSV text, independent of any parsing library.

We nonetheless encourage readers to verify against their own copy. The
repository includes a dependency-free script that does so in one command,
reading every field as text so that no type inference can mask the difference:

\begin{verbatim}
python scripts/verify_placeholder_split.py \
       ML-EdgeIIoT-dataset.csv
\end{verbatim}

The audit reduces to a single call for readers who wish to check their own
pipelines:

{\footnotesize
\begin{verbatim}
from agriedge.audit.leakage import (
    audit_columns,
    summarize,
)

reports = audit_columns(
    frame,
    categorical_columns,
    label_column="Attack_label",
)
print(summarize(reports))
\end{verbatim}
}

\section{Conclusion}

Edge-IIoTset carries a serialisation artifact that makes file provenance
linearly separable from the label. The preprocessing recipe distributed with
the dataset converts a placeholder-spelling difference between the
normal-traffic and attack-traffic build branches into features that recover the
label exactly. Four of the seven columns the recipe names separate every row of
both curated subsets on their own---157{,}800 rows in one, 2{,}219{,}201 in the
other---and under that recipe five of six standard classifiers attain exactly
1.0000 accuracy in every one of fifteen cross-validation folds and the sixth
attains 0.99998, while a 1D-CNN and a deep MLP reach 1.0000 on a single
stratified split.
The leak is not a property of one-hot encoding: label, ordinal and frequency
encoding transmit it identically.

We stop short of claiming that any particular published number is an artifact,
since preprocessing is rarely reported in sufficient detail to determine it
from outside. What we can say is stronger than a suspicion and weaker than an
indictment: the recipe the dataset's own documentation prescribes yields
perfect or all-but-perfect classification for every model we tried; the encodings
researchers substitute for it do the same; and the affected columns are the
ones an IIoT practitioner would naturally reach for. Any result on this dataset
whose preprocessing touched those columns needs re-checking before it can be
read as detection performance.

Correcting the artifact costs the strongest model about five points of macro-F1
and costs the weakest nearly a third of its score. Rebuilding the benchmark
from raw captures under uniform parsing eliminates the artifact by
construction, restores the Modbus traffic and per-device attribution that
agricultural research requires, and yields a benchmark on which no column
separates the classes above 0.0288.

On that corrected benchmark, a leave-one-device-out sweep locates where
detection actually stops working. Transfer between perception sensors is nearly
free. Transfer across the perception/actuation layer is not: withholding the
Modbus gateway drops random forest from 0.9988 to 0.5083 balanced accuracy and
logistic regression to 0.4877, straddling the trivial baseline, and inverts
the model ranking so that optimising average performance selects the model
least able to survive the shift. Combined with a 342$\times$ spread in
inference latency across models of indistinguishable random-split accuracy, the
numbers the field currently optimises say very little about deployment.

Our federated results sharpen the same point from a different direction.
Non-IID partitioning---the difficulty the federated IIoT literature most often
sets out to solve---costs at most 0.0037 macro-F1 on this benchmark, while the
uplink required to train even a 28{,}450-parameter model over LoRaWAN costs 4.6
hours of continuous transmission. The obstacle to federated intrusion detection
on a real farm is bandwidth, not statistical heterogeneity.

The corrective is inexpensive: canonicalise placeholders before encoding,
deduplicate before splitting, evaluate under held-out-domain protocols, include
a weak model as a leakage canary, and report deployment cost alongside
accuracy. We release tooling for all five.

\subsection{Future work}

Three directions follow directly. \textbf{Cross-network transfer}: establishing
whether detectors trained on AgriEdge survive a genuinely different network
requires a bridging feature space between packet-field and flow-statistical
representations, which is the principal open problem for this line of work.
\textbf{A second actuation-layer device}: our central generalisation finding
rests on a single fieldbus device; a testbed carrying several distinct
actuation protocols (Modbus alongside DNP3, BACnet, or OPC-UA) would determine
whether the boundary is a property of the layer or of this particular capture.
\textbf{Auditing other security datasets}: the failure mode requires only that
a dataset be assembled by concatenating separately-parsed sources whose
separation correlates with the label, which describes a large fraction of
intrusion-detection corpora. Applying the audit released here across the
standard benchmark suite would establish whether Edge-IIoTset is exceptional or
representative. We suspect the latter.

\section*{Data availability}

Edge-IIoTset is publicly available and free for academic use per its own
licence~\cite{ferrag2022edgeiiotset}. The \code{agriedge} package, the AgriEdge
benchmark construction pipeline, and all experiment scripts and result tables
are released under the MIT licence at
\url{https://github.com/MostafaGalal1/agriedge} and archived under the
persistent identifier \url{https://doi.org/10.5281/zenodo.21941210}, which
resolves to the most recent release. The constructed benchmark is not carried
in the repository, being 38~MB; it is published as a separate archived dataset
at \url{https://doi.org/10.5281/zenodo.21941319}, so that it can be obtained
directly without first acquiring the 10~GB source distribution. It is equally
regenerable by running \code{experiments/03\_build\_agribench.py} against a
local copy of Edge-IIoTset.

\section*{Declaration of competing interest}

The author declares no competing financial interests or personal relationships
that could have appeared to influence the work reported in this paper.

\bibliographystyle{elsarticle-num}
\bibliography{refs}

\end{document}